\documentclass[prl,twocolumn,amsmath,amssymb]{revtex4}

\usepackage{graphicx}

\begin{document}

\pagestyle{empty}
\newcommand{\isotope}[2]{\mbox{$^{#2}\mbox{#1}$}}

\noindent \textbf{Comment on ``Taleyarkhan
  {\bfseries\itshape et al.} Reply:''}

\vspace{0.3cm}

In their Reply~\cite{taleyarkhanReply} to my previous
Comment~\cite{naranjoComment}, Taleyarkhan and coauthors measure their
detectors' responses to a \isotope{Cf}{252} source, concluding that
the resulting spectra differ substantially from the cavitation-fusion
spectra published earlier in their Letter~\cite{taleyarkhan2006}.  On
the contrary, I conclude that the two data sets are qualitatively
consistent.

\begin{figure}[h]
  \centering
    \includegraphics[width=86mm]{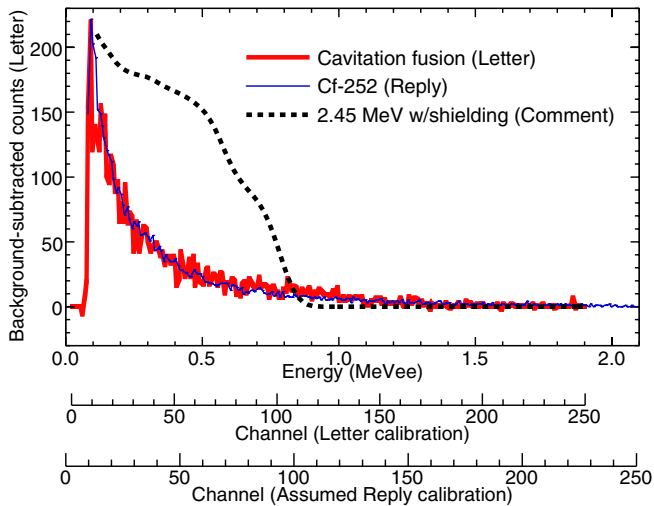}
    \caption{(color online).  The aggregate background-subtracted
cavitation-fusion proton-recoil spectrum (Fig.~12 of the Letter's
supplement~\cite{taleyarkhan2006supp}) compared with the
\isotope{Cf}{252} spectrum from the Reply.  Note that the
``PRL'' spectrum from Fig.~1(a) of the Reply is the same as the
cavitation-fusion spectrum here, though data below channel~10 were
removed from the Reply.  As discussed in the text, the Reply spectrum
is cross-calibrated to have a gain of 10\% less than the Letter
spectrum.  For qualitative comparison, the simulated DD-fusion
response from my previous Comment, with arbitrary vertical
normalization, is also shown.  See Refs.~\cite{zmeskal1990,nagele1989}
for examples of experimental DD-fusion proton-recoil spectra.}
    \label{fig:cf252_vs_dd}
\end{figure}

\paragraph{NE-213 neutron spectra.}
To compare proton-recoil spectra, their scales must first be
calibrated, typically to equivalent electron energy via $\gamma$
calibration sources.  Though the authors provided \isotope{Cs}{137}
and \isotope{Co}{60} $\gamma$ calibrations in their Letter (see
Fig.~1(a) of my previous Comment), in their Reply, they do not provide a
$\gamma$ calibration along with their \isotope{Cf}{252} spectrum.
Nevertheless, their detector's response to \isotope{Cf}{252}, and the
corresponding \isotope{Cs}{137} and \isotope{Co}{60} $\gamma$
calibrations, are given in Figs.~5(b) and 4 of Ref.~\cite{xu2005}.
Comparison of the calibrated and the uncalibrated \isotope{Cf}{252}
spectra shows that the detector's gain was approximately 10\% less in
the Reply than in the Letter.  Using this calibration for the Reply
spectrum, Fig.~\ref{fig:cf252_vs_dd} shows the Reply's
\isotope{Cf}{252} spectrum to be consistent with the Letter's
cavitation-fusion spectrum.

\begin{figure}[h]
  \centering
    \includegraphics[width=71mm]{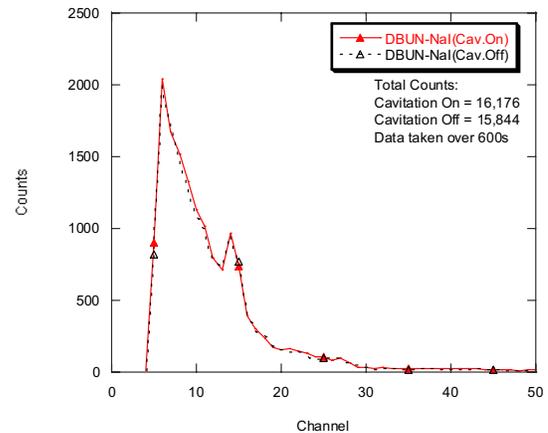}
    \caption{Fig.~15(a) of the Letter's
supplement~\cite{taleyarkhan2006supp}.}
    \label{fig:cavitation_gammas}
\end{figure}

\paragraph{NaI(Tl) gamma spectra.}
In Fig. 1(b) of the Reply, the authors compare their ``cavitation on''
$\gamma$ spectrum against an experimental \isotope{Cf}{252} $\gamma$
spectrum.  As shown in Fig.~15(a) of the Letter's
supplement~\cite{taleyarkhan2006supp} (reproduced here in
Fig.~\ref{fig:cavitation_gammas}), the ``cavitation on'' spectrum is
within approximately 2\% of the ``cavitation off'' background
spectrum.  Consequently, they are comparing the \isotope{Cf}{252}
spectrum against the natural $\gamma$ background, not the cavitation
fusion $\gamma$ signal.  For example, the peak at channel~14, also
present in their undeuterated control runs, is due to
\isotope{K}{40}'s 1.46~MeV $\gamma$, the predominant feature of
the terrestrial $\gamma$ background~\cite{petrasso1989}.  These
features do not appear in the Reply spectrum because their relatively
intense \isotope{Cf}{252} source, placed only 30~cm from the detector,
overwhelms the natural $\gamma$ background.

The appropriate comparison would be between the \isotope{Cf}{252}
$\gamma$ spectrum and the background-subtracted cavitation-fusion
$\gamma$ signal.  In this case, however, the subtracted signal is a
small fraction of the background, and the error on a channel's count
difference would be of greater magnitude than the difference itself.
For example, in channel~14, there were approximately 970~`on' counts
and 940~`off' counts, yielding a difference of $30 \pm 40$.  Such a
comparison would therefore be unfortunately inadequate to distinguish
between \isotope{Cf}{252} and DD-fusion.

In conclusion, Taleyarkhan and coauthors' cavitation-fusion spectra
are consistent with their own \isotope{Cf}{252} spectra.

\acknowledgments{I thank S.~Putterman for valuable discussions.
This work is supported by DARPA.}

\vspace{0.3cm}

\small
\noindent B. Naranjo \\
\indent UCLA Department of Physics and Astronomy \\
\indent Los Angeles, California 90095, USA \\ \\
\noindent February 1, 2007 


\end{document}